\documentclass[12pt]{article}
\pdfoutput=1        
\usepackage{amsmath,amssymb} 
\usepackage{pictfile}     
\usepackage[mathscr]{eucal}       
\usepackage{latexsym,amsfonts,amstext,verbatim}            
\usepackage{epsfig}

\numberwithin{equation}{section}

\newcommand{\nitem}{\vspace{-0.4em}\item}

\newcommand{\ccomma}{\raisebox{.5ex}{,}\relax}

\newcommand{\bsm}{\boldsymbol}

\newcommand{\rmi}{\mathrm{i}}
\newcommand{\rme}{\mathrm{e}}
\newcommand{\rmd}{\mathrm{d}}

\newcommand{\bR}{\mathbb{R}}
\newcommand{\bC}{\mathbb{C}}

\newcommand{\frD}{\mathfrak{D}}
\newcommand{\frH}{\mathfrak{H}}

\newcommand{\cC}{\mathscr{C}}

\newcommand{\cO}{\mathscr{O}}

\newcommand{\cR}{\mathscr{R}}

\newcommand{\cW}{\mathscr{W}}

\renewcommand{\[}{\hspace{0.0375em}\relax}

\begin{document}

\title{\large\bf Little-used Mathematical Structures\\[0.7mm]
in Quantum Mechanics\\[2mm] 
II. Representations of the CCR and Superseparability}

\author{ {R N Sen}\\
{\normalsize Department of Mathematics}\\[-0.7mm]
{\normalsize Ben-Gurion University of the Negev}\\[-0.7mm] 
{\normalsize 84105 Beer Sheva, Israel}\\
{\normalsize\tt E-mail: rsen@cs.bgu.ac.il}\\[1mm]
{\normalsize 22 January 2012}}
\date{}

\maketitle
\thispagestyle{empty}
\pagebreak


\begin{abstract}

\vspace{3mm} 
It often goes unnoticed that, even for a finite number of degrees of
freedom, the canonical commutation relations have many inequivalent
irreducible unitary representations; the free particle and a particle
in a box provide examples that are both simple and well-known. The
representations are unitarily inequivalent because the spectra of the
position and momentum operators are different, and spectra are
invariant under unitary transformations. The existence of these
representations can have consequences that run from the merely
unexpected to the barely conceivable. To start with, states of a
single particle that belong to inequivalent representations will
always be mutually orthogonal; they will \emph{never} interfere with
each other.  This property, called \emph{superseparability}
elsewhere, is well-defined mathematically, but has not yet been
observed.  This article suggests two single-particle interference
experiments that may reveal its existence.  The existence of
inequivalent irreducibile representations may be traced to the
existence of different self-adjoint extensions of symmetric operators
on infinite-dimensional Hilbert spaces. Analysis of the underlying
mathematics reveals that some of these extensions can be interpreted
in terms of topological, geometrical and physical quantities that can
be controlled in the laboratory.  The tests suggested are based on
these interpretations. In conclusion, it is pointed out that
mathematically rigorous many-worlds interpretations of quantum
mechanics may be possible in a framework that admits
superseparability.\pagebreak

\end{abstract}


\section{Introduction: Superseparability}\label{SEC-INTRO}

In complete contrast to the representations of Lie groups,
representations of the canonical commutation relations (hereafter
CCR) for a \emph{finite} number of degrees of freedom have attracted
little attention from physicists.\footnote{In this paper the
abbreviation CCR will always refer to a \emph{finite} number, $n$, of
degrees of freedom. The general features are independent of $n$, as
long as it is finite.} Although uncountably many
pairwise-inequivalent irreducible unitary representations (hereafter
IURs) of the CCR are routinely constructed in introductory texts on
quantum mechanics, they are not given the recognition they
deserve,\footnote{The free particle and particles in boxes of
incommensurate sizes give rise to pairwise-inequivalent
representations. This case will be discussed in
{Sec.}~\ref{SUBSEC-EX-TXTBKS}.\label{FTN-REF-1}} for the mere
existence of inequivalent IURs suffices to open the door to new and
unexpected phenomena.

In von Neumann's formulation of quantum mechanics, it is tacitly
assumed that the pure states of a single particle always belong to
a fixed irreducible unitary representation $\pi$ of the CCR. If this
assumption is dropped, then a pure state of a single particle may
have components that belong to inequivalent IURs of the CCR. This
property has been called \emph{superseparability} in \cite{SEN2010}.
In the simplest case, in which only two IURs are admitted, the
single-particle Hilbert space will be 
\begin{equation}\label{DEF-SUP-0}
\frH = \frH_1\oplus\frH_2,
\end{equation}
where $\frH_{1,2}$ carry inequivalent IURs $\pi_{1,2}$ respectively.
The generic single-particle state will be
\begin{equation}\label{DEF-SUP-1}
\Psi=\Psi_1+\Psi_2, 
\end{equation} 
where 
\begin{equation}\label{DEF-SUP-2} 
\Psi_1 = \psi_1 \oplus 0,\;\; \Psi_2 = 0 \oplus\psi_2,\quad
\psi_1\in\frH_1\;\; \text{and}\;\; \psi_2\in\frH_2. 
\quad
\end{equation} 
We would then have 
\begin{equation}\label{DEF-SUP-3}
(\Psi_1,\Psi_2) = 0\;\,\forall\;\, \psi_1\in\pi_1\; \text{and}\;
\psi_2\in\pi_2,
\end{equation}
even if $\psi_1$ and $\psi_2$ have exactly the same dependence on the
space coordinates at any given time! Note that, for (\ref{DEF-SUP-3})
to hold, it is not necessary that $\pi_1$ and $\pi_2$ be inequivalent. 

Is superseparability only a property of the mathematical formalism of
quantum mechanics, or is it also reflected in nature? This question
can be answered, if at all, only by experiment, and the aim of the
present paper is to suggest two single-particle interference
experiments that may be performable in the laboratory.\footnote{By a
single-particle interference experiment we mean one in which at most
one particle traverses the interferometer at any given time.} The
suggestions will based on an \emph{analysis of the mathematical
conditions} that give rise to inequivalent representations.

This analysis requires much more mathematical machinery than the
preceding paper \cite{LUMS-1}. To devise an experiment, one has to
identify entities that influence the phenomenon and can be controlled
in the laboratory. In the case of superseparability, some of these
entities appear to be hidden in the definitions of unbounded
self-adjoint operators on a separable Hilbert space. In infinite
dimensions the concept of self-adjointness has complexities that do
not exist on finite-dimensional vector spaces.\footnote{These
complexities cannot be handled in the Dirac formalism.} Although this
concept was analyzed by von Neumann in 1929--30 \cite{vN1929-1930}
(and expounded in considerable detail in his book \emph{Mathematische
Grundlagen der Quantenmechanik} in 1932 \cite{vN1932}), only
mathematical physicists specializing in functional-analytic methods
may be assumed to be familiar with it.  Additionally, the problem we
wish to address intertwines geometrical questions concerning the
relation between Lie groups and Lie algebras with functional-analytic
questions concerning the notion of self-adjointness of operators on a
separable Hilbert space.  The present paper, which is addressed to
experimental as well as theoretical physicists, will not assume this
mathematical background, and will begin with a concise but adequate
account of the material that will be called upon. 

Mindful of what was said above, the present paper is organized as
follows.  {Section}~\ref{SEC-HIST} goes back to where it all began,
and describes some work by Hermann Weyl in 1928 and John von Neumann
in 1929-30 which set the stage for everything that followed.  Apart
from its intrinsic interest, the historical background also reveals
the geometrical aspect of the problems that we have to address. Then
comes the functional-analytic aspect, which is the theory of
unbounded symmetric and self-adjoint operators.  A brief summary of
the material that is essential for our purposes is provided in
{Sec.}~\ref{SEC-SELF-ADJ}.  Inequivalent representations of the CCR
are discussed in {Sec.}~\ref{SEC-INEQUIV}.  The discussion, aimed at
unearthing quantities that can be controlled in the laboratory rather
than at mathematical complexities of the subject, is based on three
examples: the one mentioned briefly in footnote \ref{FTN-REF-1}, one
due to Schm\"udgen and one due to Reeh.  Conditions under which
superseparability may be revealed in one-particle interference
experiments are discussed in {Sec.}~\ref{SEC-POSSIBILITIES}. Based on
this discussion, two experiments are suggested in
{Sec.}~\ref{SEC-EXPTS}: a ``2+1-slit'' far-field interferometry
experiment, and one using Reeh's observation on the Aharonov-Bohm
effect. The concluding section discusses superseparability and the
many-worlds interpretation of quantum mechanics.


\section{Historical background}\label{SEC-HIST}

We begin by recalling two basic facts. (i) The Born-Jordan
commutation relation $[p,q] = -\rmi\,\!I$ cannot be represented by
finite-dimensional matrices if $I$ is required to be the identity
matrix. (ii) If it is represented on an infinite-dimensional Hilbert
space $\frH$ with $I$ as the identity operator, then at least one of
$p$ and $q$ must be represented by an unbounded operator (see, for
example, \cite{SEN2010}). Recall that an infinite-dimensional Hilbert
space is required to be \emph{complete}, i.e., every Cauchy sequence
has to converge, and \emph{separable}, i.e. has to have a countable
orthonormal base.\footnote{The requirement of separability,
introduced by von Neumann, has since been dropped in the mathematical
literature.} These requirements are automatically satisfied by
finite-dimensional Hilbert (or inner product) spaces. All our Hilbert
spaces will be over the complex numbers.

Unbounded operators are not defined everywhere on a Hilbert space,
and are discontinuous wherever they are defined. They give rise to
mathematical phenomena that are not encountered in the theory of
finite dimensional matrices, and it requires considerable effort to
invest with meaning even the simplest of assertions, such as
$[A,B]=0$, if $A$ and $B$ are unbounded. The basic structures of
quantum mechanics, namely matrix mechanics, wave mechanics and
transformation theory were laid down in 1925--27,\footnote{Dirac's
\emph{Principles of Quantum Mechanics} was first published in 1930,
as was Heisenberg's \emph{Physical Principles of Quantum Mechanics}
\cite{H1930}.} but unbounded \mbox{operators} began to be explored
only in 1929--1930 \cite{vN1929-1930}. In retrospect, one is
struck by the fact that transformation theory could be developed with
scant understanding of the operators that were to be transformed. By
what magic was this achieved?

In 1928, Weyl published his book \emph{Gruppentheorie und
Quantenmechanik} \cite{W1928}. In this book he replaced the canonical
commutation relations for $N$ degrees of freedom by a $2N$-parameter
Lie group, which had the CCR as its Lie algebra; in one fell swoop,
he eliminated the vexing problems associated with unbounded operators
and brought the subject under the ambit of group theory.  This group
has become known as the \emph{Weyl group}, and we shall denote it by
$\cW_N$. We shall give the argument for $N=1$; the general case
merely requires a cumbersome modification of the notation (see
\cite{W1928}, pp.\ 272--276).

Let $a, b \in \bR$ and define, formally,  
\begin{equation}\label{WEYL-GROUP-1}
u(a) = \exp\,(\rmi ap),\qquad v(b) = \exp\,(\rmi bq).
\end{equation}
From the properties of the exponential function, it follows
that
\begin{equation}\label{WEYL-GROUP-2}
u(a)u(a^{\prime}) = u(a+a^{\prime}), \qquad
v(b)v(b^{\prime}) = v(b+b^{\prime}).
\end{equation}
Set $u(0)=v(0)=\mathbf{1}$ and $u(a)^{-1} = u(-a)$, $v(b)^{-1} = v(-b)$.
Formal computation yields the result
\begin{equation}\label{WEYL-GROUP-3} 
u(a)v(b)u(a)^{-1}v(b)^{-1} =
\rme^{\rmi ab}\mathbf{1}.  
\end{equation} 
By definition, the Weyl group $\cW_1$ consists of the elements
$\{u(a), v(b) | a,b \in \bR\}$, with multiplication defined by
(\ref{WEYL-GROUP-2}) and (\ref{WEYL-GROUP-3}). The element
$\mathbf{1}$ is the identity of the group. The group $\cW_1$ is
nonabelian and noncompact, with $\bR^2$ as the group manifold, and is
a Lie group. The same is true of the Weyl group $\cW_N$ for $N$
degrees of freedom, except that its group manifold is $\bR^{2N}$.

Being noncompact, Weyl groups have no finite dimensional unitary
representations. In a unitary representation, the elements $u(a)$ and
$v(b)$ of $\cW_1$ are represented by unitary operators $U(a)$ and
$V(b)$ on the Hilbert space $\frH$, and similar statements hold for
$\cW_N$.\footnote{The definition of an infinite-dimensional unitary
representation includes a continuity condition that we have not
specified. The same condition is used in the definition of
one-parameter groups of unitaries.} A result known as Stone's theorem
asserts that a one-parameter group of unitaries $\{U(t)\}$ on a
Hilbert space has an infinitesimal generator $H$, so that $U(t) =
\exp\,(\rmi Ht)$, where $H$ is self-adjoint.\footnote{Self-adjoint
operators on infinite-dimensional Hilbert spaces will be defined
precisely in {Section}~\ref{SEC-SELF-ADJ}. The exponential
$\exp\,(\rmi At),\, t\in\bR$ of the unbounded self-adjoint operator
$A$ needs definition, but we shall content ourselves with the
statement that it turns out to have the expected properties.} It is
bounded if $\{U(t)\}$ is compact ($t\in S_1$, the circle) and
unbounded if $\{U(t)\}$ is not compact ($t \in \bR$).  A
representation of $\cW_N$ defines, uniquely, a representation of its
Lie algebra -- the CCR -- by self-adjoint operators.  In the
representation so defined, the operators $p$ and $q$ are unbounded.

In 1930 von Neumann proved that, for finite $N$, the Weyl group
$\cW_N$ has only one irreducible unitary representation
\cite{vN1930}.\footnote{A much simpler proof was given later by
Mackey \cite{GWM1968}. The reader familiar with the theory of induced
representations will recall that inequivalent irreducible
representations of the little group determine inequivalent
irreducible representations of the whole group. Mackey's proof
consisted of showing that the little group consisted of the identity
alone.} He gave the name \emph{Schr\"odinger operators} to the
representatives of the canonical variables $p_j, q_j$,
$j=1,\ldots,N$, and titled his paper `Die Eindeutigkeit der
Schr\"odingerschen Operatoren' -- Uniqueness of the Schr\"odinger
Operators -- a choice that has turned out to be misleading.  His
result has become known as `von Neumann's uniqueness theorem'
(sometimes as the Stone-von Neumann uniqueness theorem).

A Lie group defines a unique Lie algebra, but the converse is not
true. The simplest examples are the covering groups of compact
non-simply-connected Lie groups. Examples of this phenomenon that are
relevant to elementary particle physics were unearthed as early as
1962 by Michel \cite{M1964}. The canonical commutation relations are
\emph{not} abstractly equivalent to the Weyl group; as we shall see
below, the $p_j,q_k$ will not even generate a Lie group unless they
are represented by self-adjoint operators, and the requirement of
self-adjointness cannot be met even in simple physical situations
(such as spaces with boundaries, cuts or holes) in which $q_j$ is the
operator of multiplication by $x_j$ and $p_k = -\rmi\partial/\partial
x_k$. 


\section{Symmetric operators; self-adjointness}\label{SEC-SELF-ADJ}

Let $\frH$ be a Hilbert space and $A:\frH\rightarrow\frH$ an operator
on it. If there exists a positive number $K$ such that $||A\psi||
\leq K ||\psi||$ for all $\psi\in\frH$, then $A$ is said to be
\emph{bounded}. If no such $K$ exists, then $A$ is said to be
\emph{unbounded}. An unbounded operator $A$ is not defined everywhere
on $\frH$; the subset $\frD(A)\subsetneq\frH$ on which it is defined
is called the \emph{domain} of $A$. If $\frD(A)$ is not dense in
$\frH$ then $A$ is not yet mathematically manageable, and one
generally assumes that $A$ is \emph{densely defined}, i.e., $\frD(A)$
is dense in $\frH$. (The topology on $\frH$ is the metric topology
defined by the metric on $\frH$.)

In the rest of this section we shall deal only with unbounded
operators, and therefore the adjective `unbounded' will be omitted.

In operator theory, an operator $A$ is called \emph{closed} if the
set of ordered pairs $\{(\psi; A\psi) | \psi \in \frD(A)\}$ is a
closed subset of $\frH\times\frH$. An operator $A_1$ is an
\emph{extension} of $A$ if $\frD(A)\subset\frD(A_1)$ and
$A_1\psi=A\psi$ for $\psi\in\frD(A)$; one writes $A \subset A_1$.  An
operator is called \emph{closable} if it has a closed extension.
Every closable operator $A$ has a smallest closed extension, which is
denoted by $\bar{A}$.

In matrix theory, the adjoint is defined by $(T\bsm{x}, \bsm{y}) =
(\bsm{x}, T^{\star}\bsm{y})$. In operator theory, one has to take
domains into consideration. Let $\varphi,\xi\in\frH$ such that
$(A\psi,\varphi)=(\psi,\xi)$ for all $\psi\in\frD(A)$, and define
$A^{\star}$ by $A^{\star}\varphi=\xi$. Then $\frD(A^{\star})$ is
precisely the set of these $\varphi$. One can show that if $A$ is
densely defined, then $A^{\star}$ is closed. Furthermore, $A^{\star}$
is densely defined if and only if $A$ is closable, and if it is, then
$(\bar{A})^{\star} = A^{\star}$.

We now come to the key definitions. If $\mathfrak{D}(A)
\subset\mathfrak{D}(A^*)$ and $A\varphi = A^*\varphi$ for all
$\varphi\in\mathfrak{D}(A)$, then $A$ is called
\emph{symmetric}.\footnote{Von Neumann used the term
\emph{Hermitian}, but current usage seems to limit this term to
operators on finite-dimensional vector spaces.} If $\mathfrak{D}(A) =
\mathfrak{D}(A^*)$ and $A\varphi=A^*\varphi$ for all $\varphi \in
\mathfrak{D}(A)$, then $A$ is called
\emph{self-adjoint}.\footnote{Von Neumann used the term
\emph{Hermitian hypermaximal}.} Self-adjoint operators form a
subclass of symmetric operators.

A symmetric operator may have no self-adjoint extension, it may have
many self-adjoint extensions, or it may have only one. In the last
case, it is called \emph{essentially self-adjoint}. One can show that
if $A$ is essentially self-adjoint, then its closure $\bar{A}$ is
self-adjoint, i.e., $\bar{A}$ is the unique self-adjoint extension of
$A$.

The fundamental differences between symmetric and self-adjoint
operators are:

\begin{enumerate}

\item The spectrum of a self-adjoint operator is a subset of the real
line, whereas the spectrum of a symmetric operator is a subset of the
complex plane; a symmetric operator is self-adjoint if and only if
its spectrum is a subset of the real line.

\item A self-adjoint operator can be exponentiated, i.e., if $A$ is
self-adjoint then $\exp\,(\mathrm{i} tA)$ is defined for all $t \in
\mathbb{R}$; a symmetric operator which is not self-adjoint
\emph{cannot} be exponentiated.

\end{enumerate}

If $A$ and $B$ are self-adjoint, defined on a common dense domain
$\mathfrak{D}$ and commute on $\mathfrak{D}$, then $\exp\,
(\mathrm{i}aA)$ and $\exp\,(\mathrm{i}bB)$ are defined for all $a,b
\in \mathbb{R}$ and commute. However, if $A$ and $B$ are merely
essentially self-adjoint, are defined on $\mathfrak{D}$ and commute
on $\mathfrak{D}$, then $\exp\,(\mathrm{i}a\bar{A})$ and $\exp\,
(\mathrm{i}b\bar{B})$ \emph{do not necessarily commute}. This fact,
which is highly counterintuitive, was unearthed by Nelson in 1958,
and is sometimes known as the \emph{Nelson phenomenon}; for details
and references, see Reed and Simon \cite{R-S1972,R-S1975}.

We shall conclude this section with an example. The group of
isometries of $\bR^2$ consists of translations and rotations. The
group of isometries of the punctured plane $\bR^2\setminus\{O\}$ is
the group of rotations about the origin $O$. What happens to the
translation operators on $\bR^2$, namely $\exp\,(\rmi ap_x)$ and
$\exp\,(\rmi bp_y)$, $a,b \in \bR$ (where $p_x = -\rmi\partial/
{\partial x}, \; p_y = -\rmi\partial/{\partial y}$), when the origin
is excised?

The operators $\partial/{\partial x},\,\partial/{\partial y}$ are
defined on sets of differentiable functions. A function which is
differentiable on $\bR^2$ is necessarily differentiable on
$\bR^2\setminus\{O\}$, but the converse is not true; the latter has a
richer supply of differentiable functions than $\bR^2$, e.g., the
function $r^{-1}\exp\,(-r^2/2)$ (which is also square-integrable).
Excising the origin has, in this case, \emph{enlarged} the set of
differentiable functions on which $p_x$ and $p_y$ are defined. We
state without proof that this enlargement changes the \emph{spectra}
of these operators, which in turn leads to the failure of
self-adjointness and exponentiability.


\section{Inequivalent representations of the CCR}
\label{SEC-INEQUIV}

Inequivalent IURs of the relativity or internal symmetry groups used
in physics are completely classified by values of the invariants of
their Lie algebras, and these invariants can be determined in the
laboratory.  By contrast, the only invariant of the Lie algebra
defined by the CCR is the identity, and a useful classification of
IURs of the CCR is not yet known. If we want to distinguish between
inequivalent IURs in the laboratory, then, in the present state of
our knowledge, our only option is to examine the mathematical
structures that give rise to inequivalent IURs in search of clues. As
stated earlier, we shall confine our search to the analysis of
specific examples.


\subsection{An example from the textbooks}\label{SUBSEC-EX-TXTBKS}

Consider first the free (spinless) particle on the real line. Its
Hilbert space is $L^2(\bR, \rmd x)$.  The operators $p$ and $q$ are
self-adjoint and are defined on a common dense domain. Their spectra
are continuous, and fill the real line. We shall denote this
representation of the CCR by $\pi_{\bR}$.

Consider now a particle which is constrained to lie in the interval
$[0,\Lambda]\subset\bR$. Its Hilbert space is the subspace $\frH_{\sf
box}$ of $L^2([0,\Lambda], \rmd x)$ consisting of (equivalence
classes of) functions that vanish at the boundaries.  Denote the
representation of the CCR on $\frH_{\sf box}$ by $\pi_{\Lambda}$.

One knows from textbook physics that the spectrum of $p$ in
$\frH_{\sf box}$ is discrete; its eigenvalues are $n\pi/\Lambda, n =
1,2,\ldots$ (we take $\hbar=1$). This is enough to establish that the
representations $\pi_{\bR}$ and $\pi_{\Lambda}$ of the CCR for one
degree of freedom are unitarily inequivalent to each other, because
\emph{the spectrum of an operator is invariant under unitary
transformations}. Furthermore, if $\Lambda^{\prime}$ is such that
$\Lambda/\Lambda^{\prime}$ is irrational, then the representations
$\pi_{\Lambda}$ and $\pi_{\Lambda^{\prime}}$ are unitarily
inequivalent. Since there are infinitely many real numbers such that
the quotient of any two of them is irrational, this example gives us
infinitely many pairwise inequivalent IURs of the CCR for one degree
of freedom. Note that the operators $p$ on the concrete Hilbert
spaces $\frH_{\sf box}$ are determined by the boundary conditions at
the ends of the intervals $[0,\Lambda]$.

These examples extend immediately to higher dimensions.


\subsection{An example due to Schm\"udgen}\label{SUBSEC-EX-SCHMUDGEN}

The example given below is Example 1 of \S 3 in \cite{SCH1983}. We
shall omit all computations and proofs.

Let $\frH = L^2(\bR^2)$ and $\varphi \in\frH$. Let $z$ be a complex
number such that $|z|=1$ but $z\neq 1$.  Define the unitary operators
$U(t), V(s)$ by

\begin{equation}\label{SCH-REP}
\begin{array}{lcl}
(U(t)\varphi)(x,y) &=& \rme^{\rmi z x}\varphi(x, y+t),\quad t \in
\bR \\[2mm]
(V(s)\varphi)(x,y) &=& \left\{\begin{array}{rcl}
   \varphi(x+s,y) &\text{for}& y >0, x \geq 0\\
                  && \text{and}\quad y>0, x+s <
   0\\[1mm]
z\varphi(x+s, y) &\text{for} & y >0, x< 0 \\
                 &&\text{and}\quad x+s \geq
 0\\[1mm] 
\varphi(x+s,y) &\text{for} & y\leq 0, x\in\bR
\end{array}
                     \right.
\end{array}
\end{equation}
for $s>0$, and similarly for $s<0$.  The operator $V(s)$ is a
translation in the $x$-direction, with a certain modification: in
the lower half-plane (including the $x$-axis), it takes $\varphi(x,y)$
to $\varphi(x+s,y)$, but, in the open upper half-plane, it multiplies
the translated function $\varphi(x+s,y)$ by $z$ whenever the $y$-axis
is crossed.

The sets $\{U(t)|t\in\bR\}$ and $\{V(s)|s\in\bR\}$ define two
one-parameter groups. Their infinitesimal generators are given by
\begin{equation}\label{SCH-PQ}
Q = x - \rmi\frac{\partial}{\partial y}\quad\text{and}\quad P =
-\rmi\frac{\partial}{\partial x}\cdot
\end{equation}
Formally, $[Q,P]=\rmi I$. Let now $\cR(s,t) = \{(x,y)|0<x\leq s,
0<y\leq t\}$, and denote by $\chi^{\cR(s,t)}$ the characteristic
function of $\cR(s,t)$, i.e., 
$$ \chi^{\cR(s,t)} = \left\{ \begin{array}{ll}
                           1, & (x,y) \in {\cR(s,t)}\\[2mm]
                           0, & (x,y) \notin {\cR(s,t)}.
                             \end{array}
                     \right.
$$
Then, for $\varphi\in\frH$, 
\begin{equation}\label{KS-GROUP}
(I - \rme^{-\rmi ts}V(-s)U(-t)V(s)U(t))\varphi = 
(1-z)\chi^{\cR(s,t)}\varphi
\end{equation}
Since $z \neq 1$, (\ref{KS-GROUP}) shows that the $U(s)$ and $V(t)$
do not satisfy the relations (\ref{WEYL-GROUP-3}) that define the
Weyl group. The representation $\pi_S$ of the CCR for one degree of
freedom determined by (\ref{SCH-REP}) turns out to be irreducible; it
is clearly not equivalent to the representation
$\pi_{\bR}$.\footnote{In the literature, the characteristic function
of a set $S$ is generally denoted by $\chi_S$.}

There exist many examples of inequivalent irreducible representations
of the CCR $[Q,P]=\rmi I$ in which $P$ and $Q$, defined on functions
that are themselves defined on various two-dimensional spaces, have
the form (\ref{SCH-PQ}) or forms similar to it; we refer the reader
to \cite{SCH1983} for details, and for references to earlier works.
As we have not found a way to relate these $P, Q$ to the degrees of
freedom of a physical system, we shall not dwell on these
representations.


\subsection{Reeh's example}\label{SUBSEC-EX-REEH}

In 1988, Helmut Reeh showed that that the Nelson phenomenon could be
found in the Aharonov-Bohm effect \cite{R1988}.  The motion of a
spinless particle of charge $q$ in a plane perpendicular to a
magnetic flux trapped along the $z$-axis is two-dimensional.  Its
canonical operators may be written, formally, as 
\begin{equation}\label{AB-OP-1} 
\bsm{p} = -\mathrm{i}\frac{\partial}{\partial\bsm{x}} +
q\bsm{A},\;\; \bsm{q} = \text{multiplication by}\;\bsm{x}.  
\end{equation} 

\noindent
Boldface symbols denote 2-vectors in the $XY$-plane. The vector
potential $\bsm{A}$ (up to a gauge) can be written, in terms of the
magnetic flux $\Phi$, as 
\begin{equation}\label{VP-FLUX} 
\bsm{A} = \frac{\Phi}{2\pi r}\bsm{e}, 
\end{equation} 
where $r = (x^2+y^2)^{1/2}$ and $\bsm{e}$ is the unit vector at
$(x,y)$ tangent to the circle $r = \text{const}$: 
\begin{equation*} 
\bsm{e} = \left(-\frac{y}{r}\ccomma{}\; \frac{x}{r}\right)\cdot 
\end{equation*}
We shall set 
\begin{equation}\label{ALPHA} 
\alpha = \frac{q{\Phi}}{{2\pi}} 
\end{equation} 
and use (\ref{VP-FLUX}) to rewrite the quantities $\bsm{p}$ of
(\ref{AB-OP-1}) as 
\begin{equation}\label{P-A} 
\bsm{p}^{\alpha} = -\rmi\frac{\partial}{\partial\bsm{x}} +
\alpha\frac{\bsm{e}}{r}, 
\end{equation} 
where the $\alpha$-dependence of $\bsm{p}$ has been rendered explicit
on the left. The problem is to define the formal quantities
$p_x^{\alpha}$ and $p_y^{\alpha}$ in (\ref{P-A}) as operators on the
Hilbert space $L^2(\mathbb{R}^2 \setminus O) = L^2(\mathbb{R}^2)$;
excision of a single point, here the origin $O$, has no real effect
on an $L^2$-space, but -- as we have seen earlier -- changing the
domains of differentiation operators ever so slightly can have
drastic consequences.  Reeh chose, for the common domain of
$p_x^{\alpha},\, p_y^{\alpha}$, the space $\mathscr{D}(\mathbb{R}^2
\setminus O)$ of smooth functions with compact support on
$\mathbb{R}^2\setminus O$. The space $\mathscr{D}(\mathbb{R}^2
\setminus O)$ is dense in $L^2(\bR^2)$, and $p_x^{\alpha}$ and
$p_y^{\alpha}$ are operator-valued distributions on it.\footnote{For
the concept of operator-valued distributions, see \cite{SW1964}.} If
$\varphi \in \mathscr{D}(\mathbb{R}^2 \setminus O)$, then it follows
from $\text{curl}\,\bsm{A} = 0$ that $[p_x^{\alpha},\,
p_y^{\alpha}]\varphi = 0$. 

Consider now the equation
\begin{equation}\label{EIGEN}
{p}^{\alpha}_x\varphi = \left(-\rmi\frac{\partial}{\partial x} -
\alpha\frac{y}{x^2+y^2}\right)\varphi
=  \lambda\varphi.
\end{equation}
It is a linear homogeneous differential equation of the first order
which can be solved explicitly for any $\lambda\in\bC$, and the same
holds for the equation ${p}_y^{\alpha}\psi = \lambda\psi$.  The
solutions do not have compact support. By exploiting these solutions,
Reeh established the following results \cite{R1988}:

\begin{enumerate}

\renewcommand{\theenumi}{(\alph{enumi})}
\renewcommand{\labelenumi}{\theenumi}

\item The operators $p_x^{\alpha}$ and $p_y^{\alpha}$ 
are not self-adjoint; they are essentially self-adjoint.

\item Let $\bar{p}_x^{\alpha}$ and $\bar{p}_y^{\alpha}$ be their
self-adjoint extensions, and define 
\begin{equation*}
V_x^{\alpha}(a) = \exp\,(\rmi a\bar{p}_x^{\alpha}),
\qquad
V_y^{\alpha}(b) = \exp\,(\rmi b\bar{p}_y^{\alpha}).
\end{equation*} 
Reeh showed that 
\begin{equation}\label{REEH-COMM}
V_x^{\alpha}(a)V_y^{\alpha}(b)V_x^{\alpha}(a)^{-1}
V_y^{\alpha}(b)^{-1} 
 = \rme^{\rmi(\pi\alpha/2)\cdot[\epsilon(x) -
\epsilon(x+a)][\epsilon(y) - \epsilon(y-b)]}\,I, 
\end{equation}
where $I$ is the identity operator, and
\begin{equation*}
\epsilon(t) = \left\{\begin{array}{rl}
               1&\;\; t>1\\[2mm]
               -1&\;\;t<1.
              \end{array}
              \right.
\end{equation*}
\end{enumerate}
Note that the product $[\ldots][\ldots]$ in the exponent on the
right-hand side of (\ref{REEH-COMM}) can only assume the values
$0,\pm4$, so that the entire right-hand side can only assume the
values $I, \exp\,(\pm 2\pi\rmi\alpha)I$. It follows that if $\alpha$
is an integer, then the right-hand side of (\ref{REEH-COMM}) equals
the identity operator $I$ \emph{for all admissible} $x,y,a,b$, but
\emph{not} if $\alpha$ is not an integer; in this case the group
generated by the operators $\{x, y, \bar{p}_x^{\alpha},
\bar{p}_y^{\alpha}\}$ is no longer isomorphic with the Weyl group
$W_2$. Clearly, the groups generated by these operators for $\alpha =
\alpha_1, \alpha_2$ are not isomorphic with each other if
$\alpha_1-\alpha_2$ is not an integer, and therefore the
representations of the CCR (for two degrees of freedom) they define
are not unitarily equivalent. 

Suppose now that the flux $\Phi$ is trapped inside a superconducting
cylinder. It must then be an integral multiple of $\pi/e$, where $e$
is the electronic charge. Substituting $\Phi=n\pi/e$ in
(\ref{ALPHA}) and setting $q=e$, we find that $\alpha=n/2$. That is,
if the trapped flux consists of an even number of flux quanta, the
right-hand side of (\ref{REEH-COMM}) will reduce to $I$, and the
group of the CCR to $\cW_2$. This will not happen if the trapped flux
contains an odd number of flux quanta.


\section{Controlling representations of the CCR}
\label{SEC-POSSIBILITIES}

The preceding discussion has revealed several factors that appear to
affect the representation of the CCR and can be manipulated by the
experimentalist.  They are:

\begin{enumerate}

\item Topology of the single-particle configuration space (last
two paragraphs of Sec.~\ref{SEC-SELF-ADJ}).

\item Geometry of the single-particle configuration space, if the
latter is compact (Sec.~\ref{SUBSEC-EX-TXTBKS}).

\item The vector potential, if the single-particle configuration
space is not simply connected and the particle is charged.

\end{enumerate} 

In the following, we shall suggest two single-particle interference
experiments with interferometers that have two asymmetric arms. In
the first of these, the asymmetry is topological (as well as
geometrical); the exit from one arm is through a single slit, and
from the other arm through a double slit. The suggestion is based on
the assumption that the representation of the outgoing wave is
determined by the configuration space available to it, and will
therefore be different for the two arms. The second is based on the
assumption that the two arms can be magnetically shielded from each
other. Then the interferometer can be so configured that one
arm contains a solenoid whereas the other does not, and, from
Reeh's results, the outgoing waves from the different arms will
generally be in inequivalent representations. 

In these experiments, superseparability will be revealed by a major
change in the interference pattern from the one that would be
observed in its absence.


\section{Suggested experiments}\label{SEC-EXPTS}


\subsection{A 2+1-slit far-field I-D experiment}
\label{SEC-EXPT-3-SLIT}

The device used for the experiment described below will be called a
$2+1$-\emph{slit interferometer}. The interference-diffraction
(hereafter I-D) pattern produced by it will depend on the presence or
absence of superseparability. If superseparability is absent, the I-D
pattern will be the same as that produced by a triple-slit
interferometer. That being the case, in the following the phrase
`$2+1$-{slit I-D pattern}' will imply the presence of
superseparability.  We begin by describing the device.

\begin{figure}[ht]
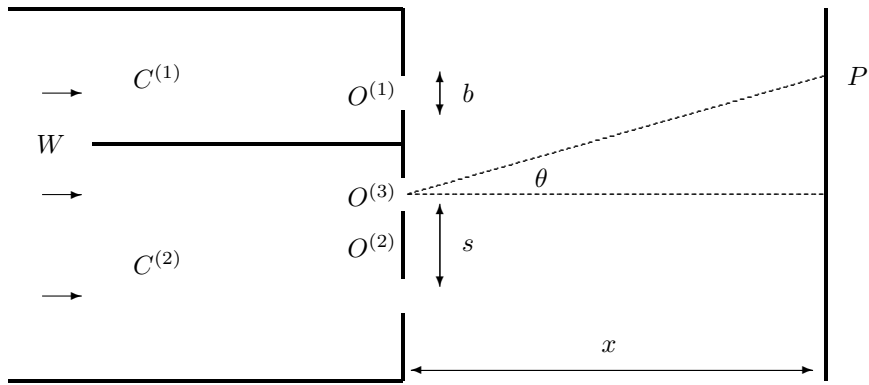


\beginpicture
\footnotesize
\setcoordinatesystem units <.75mm,.9mm>

\setplotarea x from  -90 to 90, y from -3 to 55

{\linethickness=1.25pt

\putrule from     0  0 to 0 10
\putrule from     0 15 to 0 25
\putrule from     0 30 to 0 40
\putrule from     0 45 to 0 55

\putrule from   -55 35 to 0 35

\putrule from   -70  0 to 0  0
\putrule from   -70 55 to 0 55

\putrule from    75  0 to 75 55
}
\linethickness=0.5pt

\setdashes <.5mm>

\plot 73 45  -1 27.5  74 27.5 /

\put {$W$}   at -64 35
\put {$C^{(1)}$} at -45 45
\put {$C^{(2)}$} at -45 17.5
\put {$P$}   at 79 45
\put {$\theta$} at 23 30

\put {\vector(0, -1){14.6}} [B1] at 5 45  
\put {\vector(0, 1){14.6}}  [B1] at 5 40  

\put {\vector(0, -1){28.5}} [B1] at 5 25  
\put {\vector(0, 1){28.5}}  [B1] at 5 15 

\put {\vector(1,0){150}} [B1] at 36 1.5  
\put {\vector(-1,0){150}} [B1] at 35 1.5

\put {$b$} at 10 42.5
\put {$s$} at 10 20
\put {$x$} at 35 5

\put {$O^{(3)}$} at -7 27.5
\put {$O^{(2)}$} at -7 20
\put {$O^{(1)}$} at -7 42.5

\multiput {\vector(1,0){15}}[B1] at -62 12.5  -62 27.5  -62 42.5 /

\endpicture
\caption{Interferometer for $(2+1)$-slit experiment}
\label{FIG-2+1}
\end{figure}

A cross-section of the $2+1$-slit wavefront-division interferometer
is shown in Fig.\ \ref{FIG-2+1}.  The partition $W$ separates the
chambers $C^{(1)}$ and $C^{(2)}$, which are topologically identical
except for the fact that $C^{(1)}$ has only one exit slit whereas 
$C^{(2)}$ has two.  The triple-slit interferometer $C^{(3)}$ is
exactly the same as the above, but without the partition $W$. All
devices have the same slit width $b$ and the same slit separation
$s$. 

The partition $W$ will divide the incoming wave into two.  One of
them will pass through the double slit, but not the single slit; the
other will pass through the single slit, but not the double slit.
The experiment is based on the assumption that, as a result,
\emph{the outgoing waves that emerge from $C^{(1)}$ and $C^{(2)}$
will be in inequivalent representations of the} CCR.  (In the absence
of superseparability, the presence or absence of the partition should
make little difference to the observed I-D pattern.)

The source is not shown in the figure. It is assumed that the
incoming wave, travelling in the direction of the arrows, can be
approximated by a plane wave inside the interferometer. The
wavelength will be denoted by $\lambda$. The detector is assumed to
be distant enough for observing far-field interference effects. In
this case we may base our theoretical discussion on the standard
formulae for Fraunhofer diffraction. 

We shall have to consider three I-D patterns: single-, double- and
triple-slit. The centre of a slit system (in the plane of the paper)
will be denoted by $O^{(1)}, O^{(2)}, O^{(3)}$ respectively for the
single-, double- and triple-slit systems, as shown in
Fig.~\ref{FIG-2+1}. The angle between the line of sight $OP$ and the
outward normal to the interferometer at $O$ will be denoted by
$\theta$ for all cases.

The intensity at $P$ for an $N$-slit diffraction grating is given by
the formula
\begin{equation}\label{N-SLIT-1}
I(P) =
A\left(\frac{b}{x}\right)^2\cdot\frac{\sin^2\beta}{\beta^2}\cdot
\frac{\sin^2 N\gamma}{\sin^2\gamma}\ccomma 
\end{equation}
where $A$ is a positive constant, $b$ the slit width, $x$ the
distance between the interferometer and the detector, $P$ and
$\theta$ are as shown in Fig.~\ref{FIG-2+1}, and $\beta$ and $\gamma$
are defined by 
\begin{equation}\label{BETA}
\beta = \frac{b}{\lambda}{\pi\sin \theta}\quad\text{and}\quad
\gamma = \frac{s+b}{\lambda}{\pi\sin\theta}.
\end{equation}
In the above, $s$ is the slit separation, as shown in
Fig.~\ref{FIG-2+1}. For a single slit, the \emph{interference factor}
$\sin^2 N\gamma/ \sin^2\gamma = 1$. For $N=2$ it reduces to
$\cos^2\gamma$, and for $N=3$, to $(3-4\sin^2\gamma)^2$. Formula
(\ref{N-SLIT-1}) is derived in most textbooks on physical optics, for
example \cite{JW1957}.

\begin{figure}[ht]
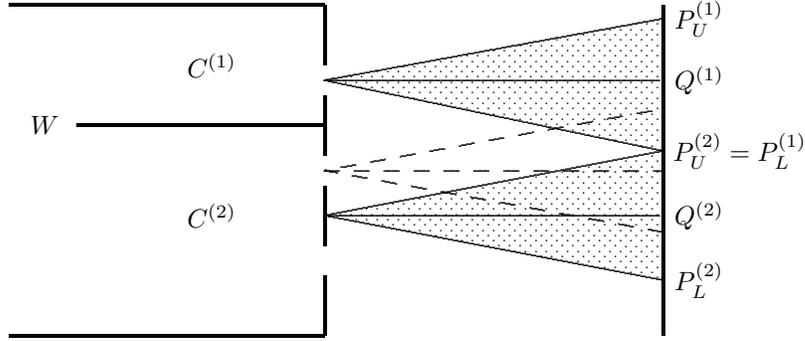


\beginpicture
\footnotesize
\setcoordinatesystem units <.6mm,0.8mm>

\setplotarea x from  -110 to 70, y from -9 to 62

\linethickness=1.25pt

\putrule from     0  0 to 0 10
\putrule from     0 15 to 0 25
\putrule from     0 30 to 0 40
\putrule from     0 45 to 0 55

\putrule from   -55 35 to 0 35
\putrule from   -70  0 to 0  0
\putrule from   -70 55 to 0 55

\putrule from    75  0 to 75 55

\plot 75 9.25   0 20  75 30.75 /  

\put {$P^{(2)}_L$} at 83 9.25
\put {$P^{(2)}_U=P^{(1)}_L$} at 92 30.75 

\plot 0 20   75 20 /

\plot 75 30.75  0 42.5  75 52.75 /

\plot 0 42.5   75 42.5 /

\put {$P^{(1)}_U$} at 83 52.75

\setdashes <2mm>

\plot 0 27.5  75 27.5 /

\plot 75 17.25  0 27.5  75 37.75 / 

\setlinear

\setshadegrid span <2pt>

\vshade 0 20 20  75 9.25 30.5 /
\vshade 0 42.5 42.5  75 30.75 52.75 /

\put {$W$}   at -62 35

\put {$C^{(2)}$} at -25 20
\put {$C^{(1)}$} at -25 45

\put {$Q^{(1)}$} at  83 42.5
\put {$Q^{(2)}$} at  83 20

\endpicture
\caption{Central maxima and first diffraction minima}
\label{FIG-DIFFR-MIN}
\end{figure}

The maxima of the \emph{diffraction factor} $\sin^2\beta/\beta^2$ in
(\ref{N-SLIT-1}) occur at $\tan \beta=\beta$, with the central
maximum at $\beta=0$. Its minima (zeroes) occur at $\beta=n\pi, n
\neq 0$, with the first minima at $\beta=\pm \pi$. Figure
\ref{FIG-DIFFR-MIN} -- not drawn to scale -- shows the positions of
the central maxima and the first diffraction minima at the detection
screen for $C^{(1)}$, $C^{(2)}$ and $C^{(3)}$.  The central maxima
are labelled by $Q$ and the first diffraction minima by $P$.  The
superscript indicates the number of slits. For the diffraction
minima, the subscripts $U,L$ denote upper and lower respectively. The
coincidence of $P^{(2)}_U$ and $P^{(1)}_L$ results from a choice of
parameters that will be explained below. The dotted lines in the
diagram indicate the diffraction cone for $C^{(3)}$, but, to avoid
crowding, the labels $P^{(3)}_L,\;P^{(3)}_U,\;Q^{(3)}$ are not shown.
Distances $d(A,B)$ between the points $A, B$ are as follows: 
\begin{equation}\label{DISTANCES}
\delta =  d(Q^{(1)}, Q^{(2)}) =
\frac32\cdot(s+b) = d(P^{(i)}_L, P^{(i)}_U), \;\; i = 1, 2, 3.
\end{equation}

From (\ref{BETA}) we see that for $N$-slit gratings ($N>1$) the slit
separation $s$ has no effect on the distances between the diffraction
minima, which are determined by the ratio $b/\lambda$ and the angle
$\theta$. What the slit separation does affect is the \emph{number of
interference maxima} between two diffraction minima.  Let $s = nb$.
Then, in the double-slit pattern, there are $n-1$ interference maxima
between the central maximum of the intensity and the first
diffraction minimum on either side of it. In the triple-slit pattern,
there are twice as many, but \emph{every other maximum
is a secondary maximum}, its intensity being roughly 11\% of the
intensities of the adjacent maxima.  In the case of superseparability
the amplitudes from $C^{(1)}$ and $C^{(2)}$ will not be added, but
the intensities will. In the absence of superseparability the I-D
pattern from $C^{(1)}+C^{(2)}$ will be the same as that from
$C^{(3)}$.

\vspace{1.7ex}   
\begin{figure}[ht]

\vspace{1.2ex}  

\noindent
\begin{minipage}[b]{.4\linewidth}
    \centering\epsfig{figure=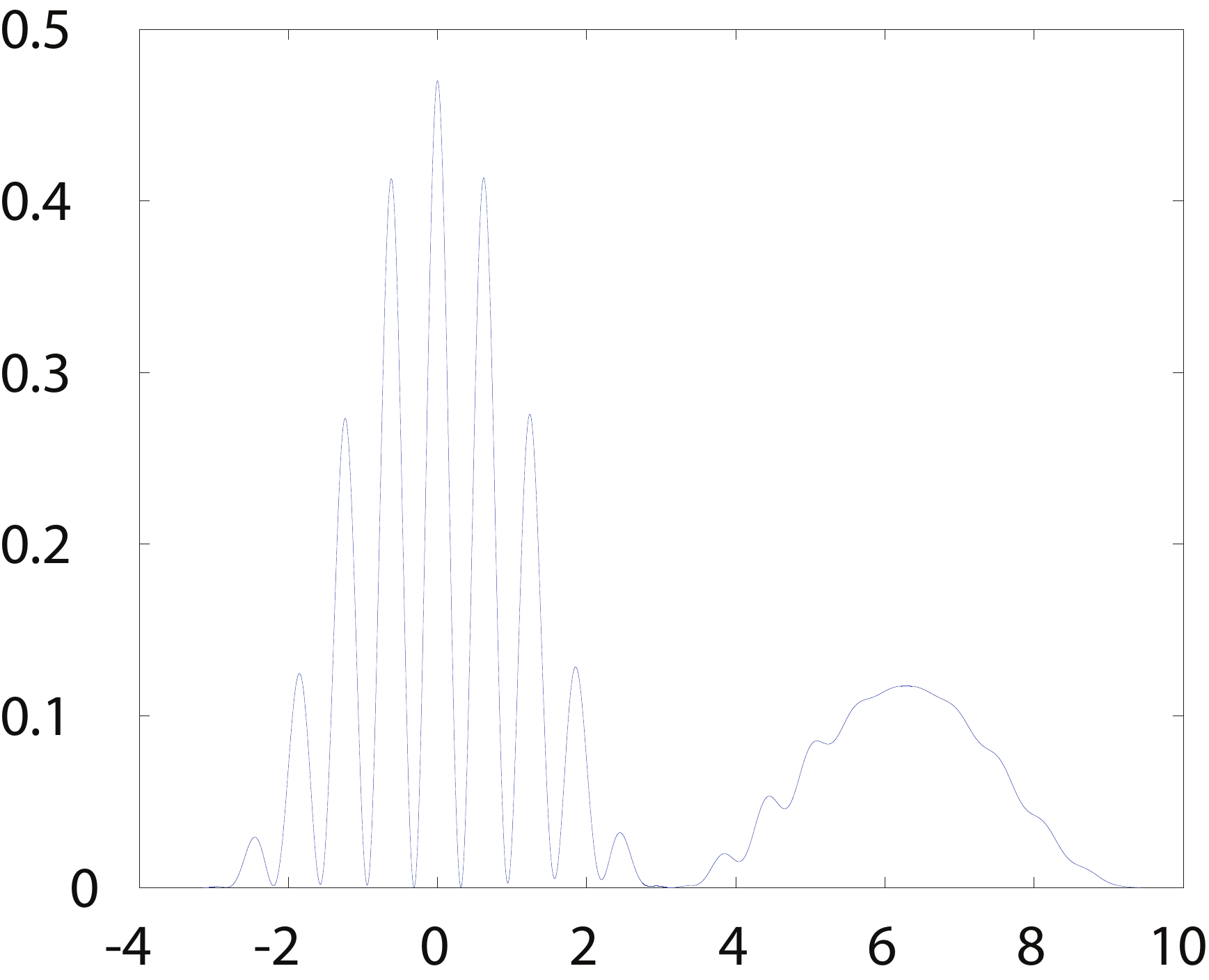,width=\linewidth}
    \caption{2+1-slit pattern} \label{FIG-2+1-SLIT}
\end{minipage}\hfill
\begin{minipage}[b]{.4\linewidth}
    \centering\epsfig{figure=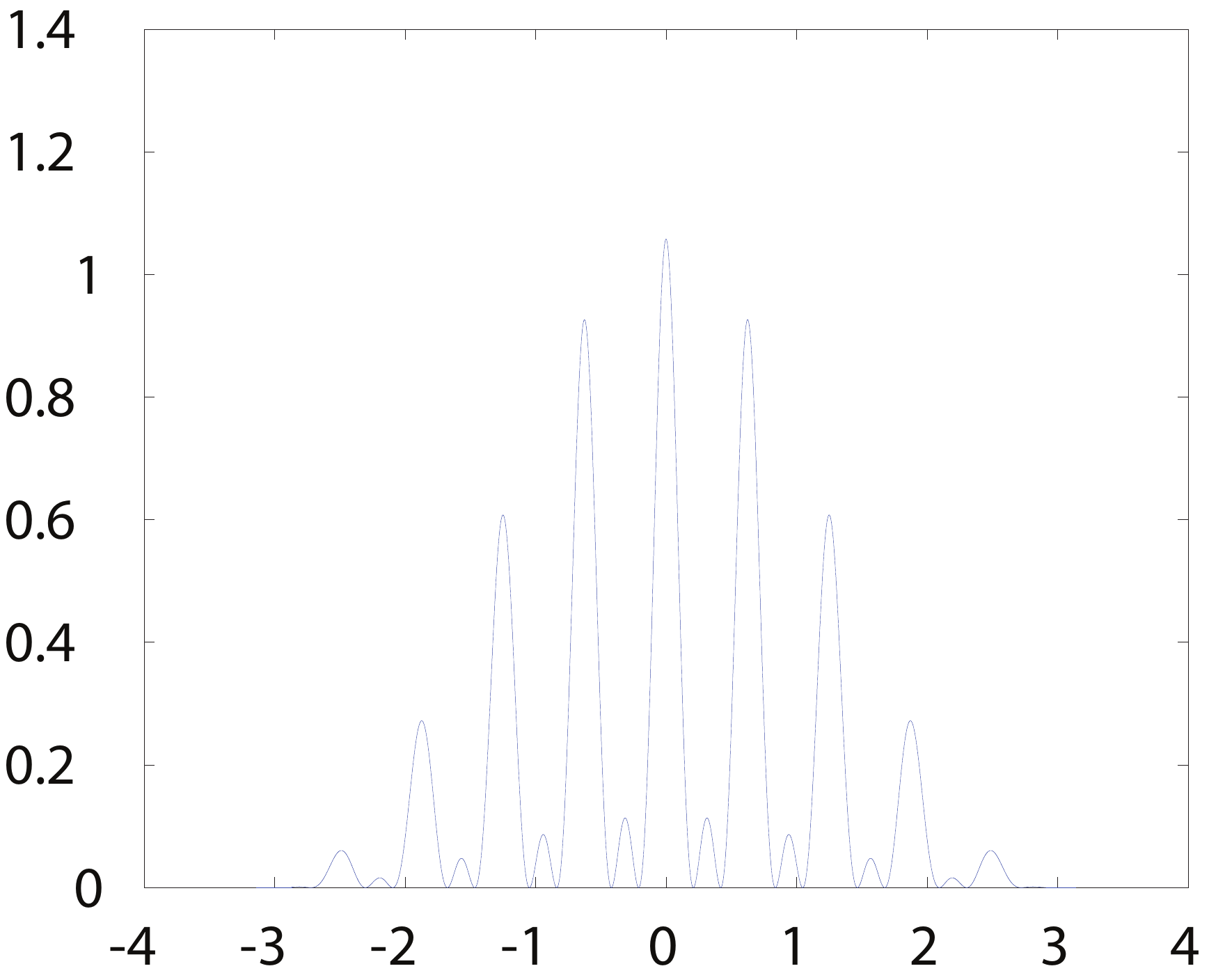,width=\linewidth}
    \caption{Triple-slit pattern} \label{FIG-3-SLIT}
\end{minipage}

\vspace{2mm}  
\end{figure}

Figures~\ref{FIG-2+1-SLIT} and \ref{FIG-3-SLIT} show the theoretical
intensity plots from the $2+1$-slit ($-\pi\leq\beta\leq 3\pi$) and
triple slit ($-\pi\leq\beta\leq\pi)$ interferometers, based on a set
of parameters that will be discussed below. The graphs depict the I-D
patterns that result when the number of counts from the single slit
is half that from the double slit, and the number of counts from the
triple slit equals that from the $2+1$-slit. 

We shall call the envelopes of the interference patterns from the
double and triple slits the \emph{diffraction} patterns. For the
single slit, there is no interference pattern, only a diffraction
pattern.  In figures~\ref{FIG-2+1-SLIT} and \ref{FIG-3-SLIT}, the
interference maxima are clearly resolved, because $s/b=5$ is small.
Were $s/b$ to be much larger, it would be impossible to resolve the
interference maxima, but \emph{the $2+1$-slit pattern would continue
to exhibit two clearly distinct diffraction maxima}, whereas the
triple-slit pattern would show only a single diffraction maximum. 

Figures~\ref{FIG-2+1-SLIT} and \ref{FIG-3-SLIT} are based on some of
the parameters of the interferometry experiment with $^{85}$Rb atoms
reported by D\"urr, Nonn and Rempe in \cite{DNR1998}.  These authors
were able to resolve, very clearly, about four peaks per millimeter
at the detector. The atoms of $^{85}$Rb were moving at about 2m/sec,
which translates to a wavelength of about $23$nm (nanometre). The
parameters assumed for the plots of Figures~\ref{FIG-2+1-SLIT} and
\ref{FIG-3-SLIT} are as follows:

\begin{enumerate}

\nitem Wavelength $\lambda=100\text{nm}=10^{-4}\text{mm}$,
corresponding to a velocity of 0.5m/sec for $^{85}$Rb atoms.

 \nitem Slit width $b=0.2\text{mm}$, so that $b/\lambda = 2\times
10^{3}$.

\nitem Slit separation $s = 1.0\text{mm}$.

\end{enumerate}

\noindent Then from $\beta=(b/\lambda)\pi\sin\theta$ it follows that 
$\beta = \pm\pi\;\text{for}\;(2\times 10^3)\sin\theta = \pm 1$. In
this case $\sin\theta\approx\theta$ is an excellent approximation,
and we find that for $\beta=\pi$, $\theta=\theta_0= 5\times 10^{-4}$.
Furthermore, we see from (\ref{DISTANCES}) and
Fig.~\ref{FIG-DIFFR-MIN} that 

$$ \frac{x}{\delta/2} =
\tan\theta_0\approx\sin\theta_0=\frac{b}{\lambda},$$ 

\noindent and from (\ref{DISTANCES}) we obtain $x=3b(s+b)/4\lambda$.
Using the chosen values of $b, s$ and $\lambda$, we find that $x=1.8$m.

The quantity $\delta=3(s+b)/2 = d(Q^{(1)}, Q^{(2)})$ is the distance
between the centres of the single-slit and double-slit systems in the
interferometer. That is, if $x=1.8$m, Fig.~\ref{FIG-DIFFR-MIN}
provides a faithful representation of the positions of the first
diffraction minima at the detection screen.
Figure~\ref{FIG-2+1-SLIT} shows the $2+1$-slit intensity pattern on
the detection screen between the points $P^{(2)}_L, P^{(1)}_U$, and
Fig.~\ref{FIG-3-SLIT} the triple-slit pattern between the points
$P^{(3)}_L, P^{(3)}_U$ of Fig.~\ref{FIG-DIFFR-MIN}.  The detector
used by D\"urr, Nonn and Rempe should be able to resolve the
interference maxima both in the double-slit and the triple-slit
patterns.


\subsubsection{Questions of feasibility}

The experiment suggested above can, in principle, be carried out
with photons, electrons, neutrons, atoms of different elements and
even fullerene molecules -- practically anything that has been used
in an interferometric experiment. The major constraints appear to be:

\begin{enumerate}

\nitem Fabricating the $2+1$-slit interferometer. The slit separation
$s$ places an upper limit on the thickness of the partitioning wall
$W$. The value of $s$ chosen above was $1$mm, or, or $1000$ microns.
The thickness of household aluminium foil is 8--25 microns, so that
there is room for improvement here. On the other hand, the distance
$\delta$ between the diffraction peaks is $\delta=3(s+b)/2$, so that
decreasing $s$ will demand a proportional increase in detector
sensetivity.

\nitem Monochromaticity of the incoming wave. This may be the dominant
constraint. In a demonstration experiment with a sodium vapour lamp
and a diffraction grating, $b/\lambda$ may be about $2\times10^{-3}$,
or even less. By contrast, with the parameters chosen above,
$b/\lambda=2\times10^3$, a difference of six orders of magnitude. In
order to observe the interference effects shown in
Fig.~\ref{FIG-2+1-SLIT} and Fig.~\ref{FIG-3-SLIT}, the incoming beam
will probably have to be monochromatic over the entire slit system.
The magneto-optical trap used by D\"urr, Nonn and Rempe as their
source may have to be augmented by a suitable monochromator, which
will result in a loss of beam intensity.

\nitem Small wavelength. The wavelength $\lambda$ may be increased by
using ultracold atoms, at the cost of reducing the count rate.  Using
lighter atoms such as $^{12}$C rather than $^{85}$Rb would produce a
seven-fold increase in the wavelength for the same count rate.  Apart
from affecting the count rate, increasing the wavelength has the same
effect as decreasing the slit width. For the same speed, the
wavelength of an electron will be $\approx1.56\times 10^{4}$ times
the wavelength of an $^{85}$Rb atom.

 \nitem Detector resolution. A tenfold increase in detector
resolution will allow $s$ and $b$ to be reduced by factors of 10. For
the same $\lambda$, $\theta_0$ will increase to $5\times 10^{-3}$,
$\sin\theta_0 \approx \theta_0$ will still be true, and $x$ will
be reduced by a factor of 10, to 18cm. 

\end{enumerate}

To sum up, it would appear that an experiment such as the one
suggested above is not unfeasible.  An experiment with photons may be
the easiest, but for the mathematically-minded theorist photons may
be the hardest objects to understand. 

Near-field effects depending on Fresnel diffraction or Talbot's bands
may offer other possibilities for the observation of
superseparability, but they require separate analysis, and have not
been considered here.


\subsection{An experiment based on Reeh's example}
\label{SEC-EXPT-REEH}

\begin{figure}[ht]
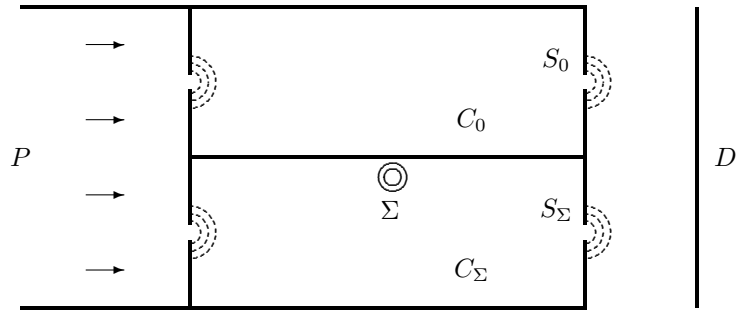


\beginpicture

\footnotesize

\setcoordinatesystem units <.75mm,1mm>

\setplotarea x from  -130 to 25, y from -8 to 48

\linethickness=1.5pt

\putrule from    -70 20   to  0 20


\putrule from    -100 0  to  0 0

\putrule from    -100 40 to  0 40


\putrule from    -70 -.3  to -70 9
\putrule from    -70 11 to -70 29
\putrule from    -70 31 to -70 40.3 


\putrule from      0 -.3  to   0 9
\putrule from      0 11 to   0 29
\putrule from      0 31 to   0 40.3 

\putrule from   20 0 to  20 40  

\circulararc 360 degrees from -36.5 17.35 center at -34 17.35
\circulararc 360 degrees from -35.5 17.35 center at -34 17.35

\linethickness=0.5pt

\setdashes <0.5mm>

\circulararc -180 degrees from 0 11.5 center at 0 10
\circulararc -180 degrees from 0 12.5 center at 0 10
\circulararc -180 degrees from 0 13.5 center at 0 10

\circulararc -180 degrees from -70 11.5 center at -70 10
\circulararc -180 degrees from -70 12.5 center at -70 10
\circulararc -180 degrees from -70 13.5 center at -70 10

\circulararc -180 degrees from 0 31.5 center at 0 30
\circulararc -180 degrees from 0 32.5 center at 0 30
\circulararc -180 degrees from 0 33.5 center at 0 30

\circulararc -180 degrees from -70 31.5 center at -70 30
\circulararc -180 degrees from -70 32.5 center at -70 30
\circulararc -180 degrees from -70 33.5 center at -70 30

\multiput {\vector(1, 0){14.6}} [B1] at -85 5  -85 15  -85 25
           -85 35  /

\put {$P$} at -100 20
\put {$\Sigma$} at -34.5 13

\put {$S_{\Sigma}$} at  -5 13
\put {$S_0$}        at  -5 33

\put {$C_{\Sigma}$} at -20 5
\put {$C_0$} at -20 25

\put {$D$} at 25 20

\endpicture
\caption{Interference experiment based on Reeh's example}
\label{FIG-INT-REEH}
\end{figure}

The experiment suggested below, based on Reeh's example, is a
single-particle interference experiment with charged particles
(typically electrons).  The scheme of the experiment is shown in
Fig.~\ref{FIG-INT-REEH}. The source $P$ is far away. $C_0$ and
$C_{\Sigma}$ are two chambers separated by a wall; they form the arms
of the interferometer. $\Sigma$ is a long thin solenoid,
perpendicular to the plane of the paper. The current through it is
controlled by the experimentalist. The configuration should be such
that clearly discernible interference fringes build up at the
detector $D$ when there is no current through the solenoid.  However,
when a current is flowing through the solenoid then, if the
phenomenon of superseparability exists, the interference pattern
should disappear, to be replaced by two separate diffraction peaks.

From the remarks at the end of Sec.~\ref{SUBSEC-EX-REEH}, one sees
that the experiment can also be performed with a superconducting
solenoid, provided that the trapped flux is an odd multiple of the
flux quantum.

The experiment is based on the assumption that waves emerging from
the slits $S_{\Sigma}$ and $S_0$ belong to inequivalent irreducible
representations of the CCR.  The solenoid $\Sigma$ is assumed to
control the representation of the wave emerging from the slit
$S_{\Sigma}$, and, at the same time, to have no effect on the
representation of the one emerging from the slit $S_0$.  Therefore
the first question we have to ask is the following: what are the
conditions under which the latter assumption may be valid?

The assumption will be valid if the vector potential due to the
solenoid is `confined' to the chamber $C_{\Sigma}$. The fact that a
material can confine magnetic \emph{fields} does not imply that it
can also confine vector potentials that cannot be gauged away. 

Suppose now that an experiment is performed with the apparatus shown
in Fig.~\ref{FIG-INT-REEH} (in which the path taken by the wave forms
a loop around the solenoid). Then, if the wall separating the
chambers $C_0$ and $C_{\Sigma}$ were not present, the experiment
would simply be one to detect the magnetic Aharonov-Bohm effect. If
the configuration shown in Fig.~\ref{FIG-INT-REEH} does confine the
vector potential due to the solenoid to $C_{\Sigma}$, then the wave
through $C_0$ would not suffer a phase shift. Therefore, if the
influence of stray fields is small, this experiment can have three
results:

\begin{enumerate}

\nitem Superseparability is detected.

\nitem Superseparability is not detected; the interference pattern
shows the fringe shift to be expected from the Aharonov-Bohm effect.

\nitem Superseparability is not detected; the interference pattern
shows \emph{half} the fringe shift to be expected from the
Aharonov-Bohm effect.

\end{enumerate}

If superseparability is detected, it would imply that the material
which confines the magnetic field also confines the vector potential.
If superseparability is not detected and the full Aharonov-Bohm phase
shift is observed, it would imply that the vector potential cannot be
confined by the material that confines the fields -- the experiment
is \emph{incapable} of detecting superseparability.  But, if a fringe
shift is observed which is only half of what would be expected from
the Aharonov-Bohm effect, it would imply that (a)~the material
\emph{does} confine the vector potential, and therefore (b)~the
phenomenon of superseparability does not exist under the given
conditions.

Since the chambers $C_0$ and $C_{\Sigma}$ cannot be completely closed
-- each will have an entrance and an exit -- the question of stray
fields, which was raised to cast doubts on the results of early
experiments on the Aharonov-Bohm effect, may be raised again. In
retrospect one sees that the effect was clearly observed, most
particularly by M\"ollensted and Bayh, well before the definitive
experiment by Tonomura. For details, the reader is referred to the
monograph by Peshkin and Tonomura \cite{P-T1989}. We therefore
believe that in the experiment suggested above, the effect of stray
fields will be negligible. However, in view of the smallness of the
flux quantum, it may be necessary to shield the apparatus from the
earth's magnetic field.


\section{Superseparability and many worlds}\label{SEC-CONCL}

If single-particle states can exist in inequivalent IURs of the CCR,
it would be natural to ask how states in inequivalent IURs interact
with each other. One unexpected possibility will be briefly discussed
in the following. 

Consider the single-particle Hilbert space $\frH=\frH_1\oplus\frH_2$
of Sec.~\ref{SEC-INTRO}. Denote by $\cO_{\sf max}$ the set of all
self-adjoint operators on $\frH$. This set contains operators $O$
such that the matrix element $(\Psi_1, O\Psi_2)\neq 0$, where
$\Psi_{1,2}$ are defined by (\ref{DEF-SUP-2}); the particle appears
to be changing representations, from $\pi_1$ to $\pi_2$ and back, due
to \emph{self-interaction}.  It is evident that the subset $\cO_{\sf
min} \subset \cO_{\sf max}$ defined by $\cO_{\sf min}=\{A_1\oplus
A_2\}$, where $A_{1,2}$ are self-adjoint operators on $\frH_{1,2}$
respectively, does not contain any self-interaction operator.
Therefore excluding such self-interactions is equivalent to taking
out of consideration the self-adjoint operators in the difference set
$\cO_{\sf max} \setminus \cO_{\sf min}$.

Consider now a two-particle system in the presence of
superseparability.  Let $\frH=\frH_1\otimes\frH_2$ be a Hilbert space
which carries the representation $\pi_1\otimes\pi_2$, where
$\pi_{1,2}$ are inequivalent IURs of the CCR. One would like to
identify the set of admissible interaction operators on $\frH$. At
one extreme is the set $\cC_{\sf max}$ of all self-adjoint operators
on $\frH$. In the absence of restrictions other than self-adjointness
on a Hilbert space carrying a single IUR, the other extreme would
appear to be the set 
\begin{equation}\label{ADMISS-OP} 
\cC_{\sf min} = \{A_1\otimes A_2\,|\,A_k\; \text{self-adjoint on}\;
\frH_k,\; k=1,2\}. 
\end{equation} 

The set (\ref{ADMISS-OP}) \emph{clearly excludes all operators that
can mediate a quantum-mechanical interaction between particles
belonging to \emph{IUR}s $\pi_1$ and $\pi_2$ that are not equivalent
to each other.} It is as if inequivalent IURs describe different
quantum-mechanical worlds. Note that the two particles can still
interact classically with each other! Note also that the
inequivalence of $\pi_1$ and $\pi_2$ is crucial; if $\pi_1=\pi_2$,
then the admissible set \emph{cannot} be $\cC_{\sf min}$; it has to
be a set $\cC$ such that $\cC_{\sf max} \supset \cC \supsetneq
\cC_{\sf min}$.

We shall not speculate any further, except for pointing out one
intriguing possibility: the existence of inequivalent irreducible
representations of the CCR may allow a mathematically rigorous
formulation of the many-worlds interpretation of quantum mechanics.

\vspace{2ex}\noindent {\bf Acknowledgements} The author would like to
thank Professor H Roos for his comments on an earlier version of this
article, and Dr M Goldberg for preparing Figs.~\ref{FIG-2+1-SLIT} and
\ref{FIG-3-SLIT}.


\begin{thebibliography}{99}

\footnotesize

\frenchspacing



\bibitem{D1958} Dirac, P A M (1958). \emph{The Principles of Quantum
Mechanics}, Fourth edition (Oxford: The Clarendon Press). First
edition, 1930.

\bibitem{DNR1998} D\"urr, S, Nonn, T and Rempe, G (1998). Origin of
quantum-mechanical complementarity probed by a `which-way' experiment
in an atom interferometer, \emph{Nature}, {\bf 395}, 33--37.

\bibitem{H1930} Heisenberg, W (1930). \emph{The Physical Principles of
Quantum Theory} (Chicago: University of Chicago Press). English
translation of \emph{Die physikalischen Prinzipien der
Quantenmechanik}, 1930 (Leipzig: Hirzel-Verlag). 

\bibitem{JW1957} Jenkins, F A and White, H E (1957). \emph{Principles
of Optics}, 3rd ed (New York: McGraw-Hill).


\bibitem{GWM1968} Mackey, G W (1968). \emph{Induced Representations
of Groups and Quantum Mechanics} (New York-Torino: W A
Benjamin-Editore Boringhieri).

\bibitem{M1964}  Michel, L (1964). Invariance in quantum mechanics
and group extension, in \emph{Group-Theoretical Concepts and Methods
in Elementary Particle Physics}, Lectures of the Istanbul Summer
School of Theoretical Physics 1962, Ed F G\"ursey (New York:
Gordon and Breach) pp~135-200.

\bibitem{P-T1989} Peshkin, M and Tonomura, A (1989). \emph{The
Aharonov-Bohm Effect} (Berlin: Springer-Verlag) Lecture Notes in
Physics Vol~340. 


\bibitem{R-S1972} Reed, M and Simon, B (1972). \emph{Methods of
Modern Mathematical Physics} I Functional Analysis (New York:
Academic Press).

\bibitem{R-S1975} Reed, M and Simon, B (1975). \emph{Methods of
Modern Mathematical Physics}, II Fourier Analysis, Self-Adjointness
(New York: Academic Press).

\bibitem{R1988} Reeh, H (1988). A remark concerning canonical
commutation relations, \emph{J Math Phys} {\bf 29} 1535-1536.

\bibitem{SCH1983} Schm\"udgen, K (1983). On the Heisenberg
commutation relation II, \emph{Publ. Res. Inst. Math. Sci. Kyoto} {\bf
19} 601-671.

\bibitem{SEN2010} Sen, R N (2010). \emph{Causality, Measurement Theory
and the Differentiable Structure of Space-Time} (Cambridge: Cambridge
University Press).

\bibitem{LUMS-1} Sen, R N (201?). Little-used Mathematical Structures
in Quantum Mechanics, I. Galilei Invariance and the \emph{welcher
Weg} Problem (preceding paper).


\bibitem{SW1964} Streater, R F and Wightman, A S (1964). \emph{PCT,
Spin and Statistics, and All That} (New York: W A Benjamin).


\bibitem{vN1929-1930} von Neumann, J (1929-1930). {Allgemeine
Eigenwerttheorie Hermitescher Funktional\-operatoren}, \emph{Math
Ann} {\bf 102} 49--131.

\bibitem{vN1930} von Neumann, J (1930). Die Eindeutigkeit der
Schr\"odingerschen Operatoren, \emph{Math Ann} {\bf 104} 570-578.

\bibitem{vN1932} von, Neumann, J (1932). \emph{Mathematische Grundlagen
der Quantenmechanik} (Berlin: Julius Springer). English translation
(revised by the author): \emph{Mathematical Foundations of Quantum
Mechanics}, 1955 (Princeton: Princeton University Press). 

\bibitem{W1928} Weyl, H (1950?). \emph{The Theory of Groups and Quantum
Mechanics} (New York: Dover Press). Translated from the 2nd German
edition. [The first German edition of \emph{Gruppentheorie und
Quantenmechanik} was published in 1928.]


\end{thebibliography}
\end{document}